\documentclass[conference,10pt]{IEEEtran}
\usepackage[cmex10]{amsmath}
\usepackage{cite, graphicx, amssymb, amsfonts} 
\usepackage[tight, footnotesize]{subfigure}
\usepackage[ruled]{algorithm2e}
\newtheorem{lemma}{Lemma}

\begin{document}
%
% paper title
\title{Joint Coding and Scheduling Optimization in Wireless Systems with Varying
  Delay Sensitivities}
%
%
% author names and IEEE memberships
% note positions of commas and nonbreaking spaces ( ~ ) LaTeX will not break
% a structure at a ~ so this keeps an author's name from being broken across
% two lines.
% use \thanks{} to gain access to the first footnote area
% a separate \thanks must be used for each paragraph as LaTeX2e's \thanks
% was not built to handle multiple paragraphs
\author{\IEEEauthorblockN{Weifei Zeng}
\IEEEauthorblockA{Research Laboratory of Electronics,\\Massachusetts Institute
of Technology,\\
Cambridge, MA, USA, \\
Email: weifei@mit.edu}
\and
\IEEEauthorblockN{Chris T. K. Ng}
\IEEEauthorblockA{Bell Laboratories,\\
Alcatel-Lucent,\\
Holmdel, NJ, USA,\\
Email: chris.ng@alcatel-lucent.com}
\and
\IEEEauthorblockN{Muriel M\'edard}
\IEEEauthorblockA{Research Laboratory of Electronics,\\Massachusetts Institute
of Technology,\\
Cambridge, MA, USA, \\
Email: medard@mit.edu}}

% make the title area
\maketitle

\begin{abstract}
Throughput and per-packet delay can present strong trade-offs that are important
in the cases of delay sensitive applications. We investigate such trade-offs
using a random linear network coding scheme for one or more receivers in single
hop wireless packet erasure broadcast channels. We capture the delay
sensitivities across different types of network applications using a class of
delay metrics based on the norms of packet arrival times. With these delay
metrics, we establish a unified framework to characterize the rate and delay
requirements of applications and optimize system parameters. In the single
receiver case, we demonstrate the trade-off between average packet delay, which
we view as the inverse of throughput, and maximum ordered inter-arrival delay
for various system parameters. For a single broadcast channel with multiple
receivers having different delay constraints and feedback delays, we jointly
optimize the coding parameters and time-division scheduling parameters at the
transmitters. We formulate the optimization problem as a Generalized Geometric
Program (GGP). This approach allows the transmitters to adjust adaptively the
coding and scheduling parameters for efficient allocation of network resources
under varying delay constraints. In the case where the receivers are served by
multiple non-interfering wireless broadcast channels, the same optimization
problem is formulated as a Signomial Program, which is NP-hard in general. We
provide approximation methods using successive formulation of geometric programs
and show the convergence of approximations.
\end{abstract}


\section{Introduction}
\label{sec:intro}
The growing diversity of network applications, protocols and architectures poses
new problems related to the fundamental trade-offs between throughput and delay
in communications.
% With the existence of the fundamental trade-offs between throughput and delay in
% communication systems, it is important to capture precisely the delay-throughput
% characteristics of applications for systems to efficiently allocate resources
% and enhance the QoE of users. However, given the growing diversity of network
% applications, protocols and architectures, it becomes increasingly challenging
% to understand the different needs of applications in terms of delay and rate and
% to design systems to accommodate them.
For instance, applications like file downloading or FTP protocols aim solely to
maximize transmission rate and to minimize the overall completion time. On the
other hand, applications such as real-time video conferencing are highly
sensitive to delay of any consecutive packets. Failure to meet continuous
delivery deadlines in stream of packets quickly deteriorates the Quality of
user Experience (QoE). The two extremes in delay sensitivities by no means represent
all types of applications. Progressive downloading video, for example, would be
more delay sensitive than file downloading, but less sensitive than real-time
video streaming, since the receiver has buffered sufficient content. 

% summarize paper content and related works
In this paper, we develop a unified framework to study rate and delay trade-offs
of coding and scheduling schemes and to optimize their performance for
applications with different delay sensitivities. We use a class of delay metrics
based on the $\ell_p$-norms of the packet arrivals times to represent delay-rate
characteristics and requirements of applications. At one extreme, the delay
metric could capture the average delay and thus the rate of transmission.  At
the other extreme, the metric measures the maximum ordered inter-arrival delay.
Based on the delay metrics, we look to optimize coding and scheduling parameters
in a networking system, where various devices with different delay requirements
are served by single-hop wireless erasure broadcast channels, each associated
with an access point (AP).

The coding scheme in this paper is a variation of the generation-based random
linear network coding, presented in \cite{ho_random_2006} and
\cite{chou_practical_2003}. Specifically, the sender maintains a \textit{coding
bucket} for each receivers. When a transmitter is ready to send a packet to some
receiver, it reads the all the packet in the coding bucket for the receiver and
produces an encoded packet by forming a random linear combination of all the
packets in the coding bucket. The encoded packet is then broadcasted to all the
receivers. Once a receiver collects enough packets to decode all packets in the
coding bucket through Gaussian elimination, it uses a separate feedback channel
to send an ACK message back to the sender. The sender always receives the ACK
message after a certain delay. It then purges all the packets in the coding
bucket and moves new packets into the bucket.  The respective delay constraints
of the receivers are known to the sender, who determines adaptively the number
of packets to put in the coding buckets for each receiver, by solving
system-wise optimization problems.  A precise description of the transmission
scheme is given in Section \ref{sec:delay}. The coding buckets act as the Head
of Line (HOL) generations in the most generation based scheme. However, unlike
most generation-based schemes, packets are not partitioned prior to transmission
and  the bucket sizes in our scheme may vary over time and across different
receivers, depending on each receiver's changing delay constraints.  The coding
parameters are optimized jointly with time division resource allocation
parameters to exploit the trade-offs between rate and delay. We first illustrate
the trade-offs in the case of point-to-point erasure channels.  Then, in the
case of multiple receivers with one AP, we formulate the delay constrained
optimization problem as a Generalized Geometric Program, which can be very
efficiently solved. We compare the solutions with fixed generation size schemes
for specific examples.  Finally, in the case of multiple APs with
non-interfering erasure broadcast channels, we formulate the problem as a
Signomial Program and provide methods to approximate this non-convex
optimization with successive GPs.

There exists a significant amount of related literature and we shall only
examine a incomplete set of relevant ones.Previous work by Walsh \textit{et al.}
\cite{maclaren_walsh_optimal_2009} considers the rate and delay trade-off in
multipath network coded networks, while \cite{ying_li_optimal_2009} studies the
related issue of rate-reliability and delay trade-off by constructing various
network utility maximization (NUM) problems.  The concept of network coding is
introduced in \cite{ahlswede_network_2000} and linear network coding is
extensively studied in \cite{li_linear_2003} and \cite{koetter_algebraic_2003}.
Other typical rateless codes that are asymptotically optimal for erasure
channels are seen in
\cite{luby_lt_2002}\cite{shokrollahi_raptor_2006}\cite{mackay_fountain_2005}.
However, unlike linear network codes which allow intermediate nodes to recode
packets, the class of fountain codes are generally only used for one-hop
communication systems, as the packets can not be recoded due to stringent packet
degree distribution requirements. In our system, the delay constraints make it
difficult to apply fountain codes efficiently, as the asymptotic optimality is
only achieved with coding over relatively large number of packets. On the other
hand, we have feedback which will allow us to dynamically change the coding
parameters. The network coding gain in overall delay of file downloading with
multicast over packet erasure broadcast channel is characterized in
\cite{eryilmaz_delay_2008} and \cite{dong_nguyen_wireless_2009}. With the use of
similar linear network codes, broadcast coding schemes based on perfect
immediate feedback are proposed and their delay characteristics are analyzed in
\cite{sundararajan_feedback-based_2009}
\cite{keller_online_2008}\cite{sadeghi_adaptive_2009}. An analysis of random
linear codes with finite block size is given in \cite{lun_analysis_2006}.  

% Paper Organization
The remainder of the paper is organized as follows. Section \ref{sec:delay}
introduces our model, the code and delay metrics, as well as how the metrics apply
specifically to the coding scheme. Section \ref{sec:gp} gives a concise primer
on Geometric Programming (GP), which is the basic tool for solving our
optimization program. 
Section \ref{sec:singleCase} considers a single wireless broadcast channel with
packet erasures. We construct a joint optimization program, which is solved using GP techniques. Furthermore
we illustrate the delay and throughput trade-offs with different system
parameters and compare the solutions with fixed generation size schemes. Section
\ref{sec:multiCase} extends on the results to multiple non-interfering wireless
channels. We provide approximation algorithms to the non-convex
optimization problem in this case. Section \ref{sec:conclusion} concludes the
paper.

\section{Delay Metrics and Coding}
\label{sec:delay}
\subsection{Adaptive Linear Coding Scheme}
\begin{figure*}[bt]
  \begin{center}
    \subfigure[Adaptive linear network coding based transmission model]{
    \includegraphics[scale=0.55]{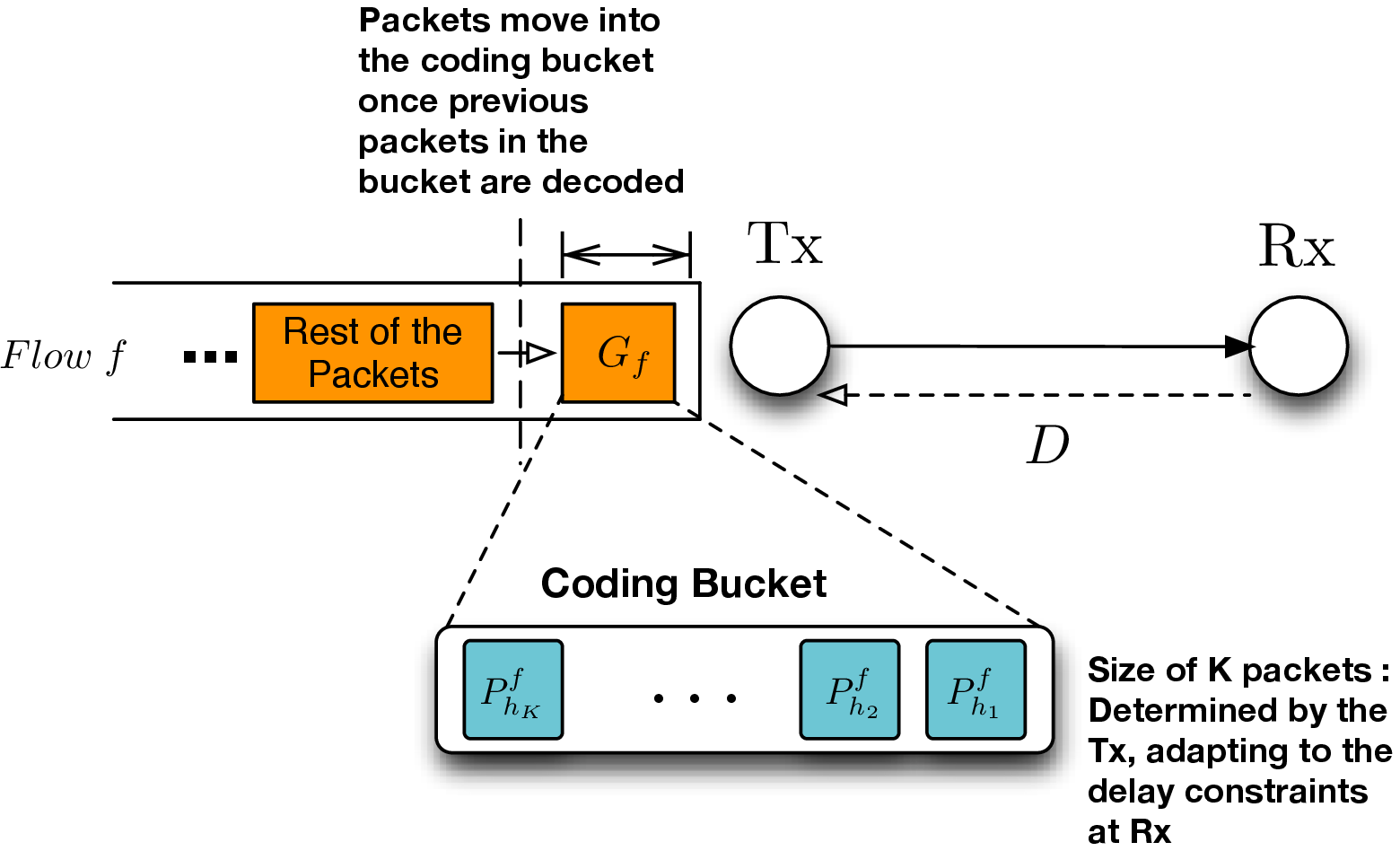}
    \label{fig:codingScheme1}
    }
    \subfigure[A session with varying coding bucket size]{
    \includegraphics[scale=0.55]{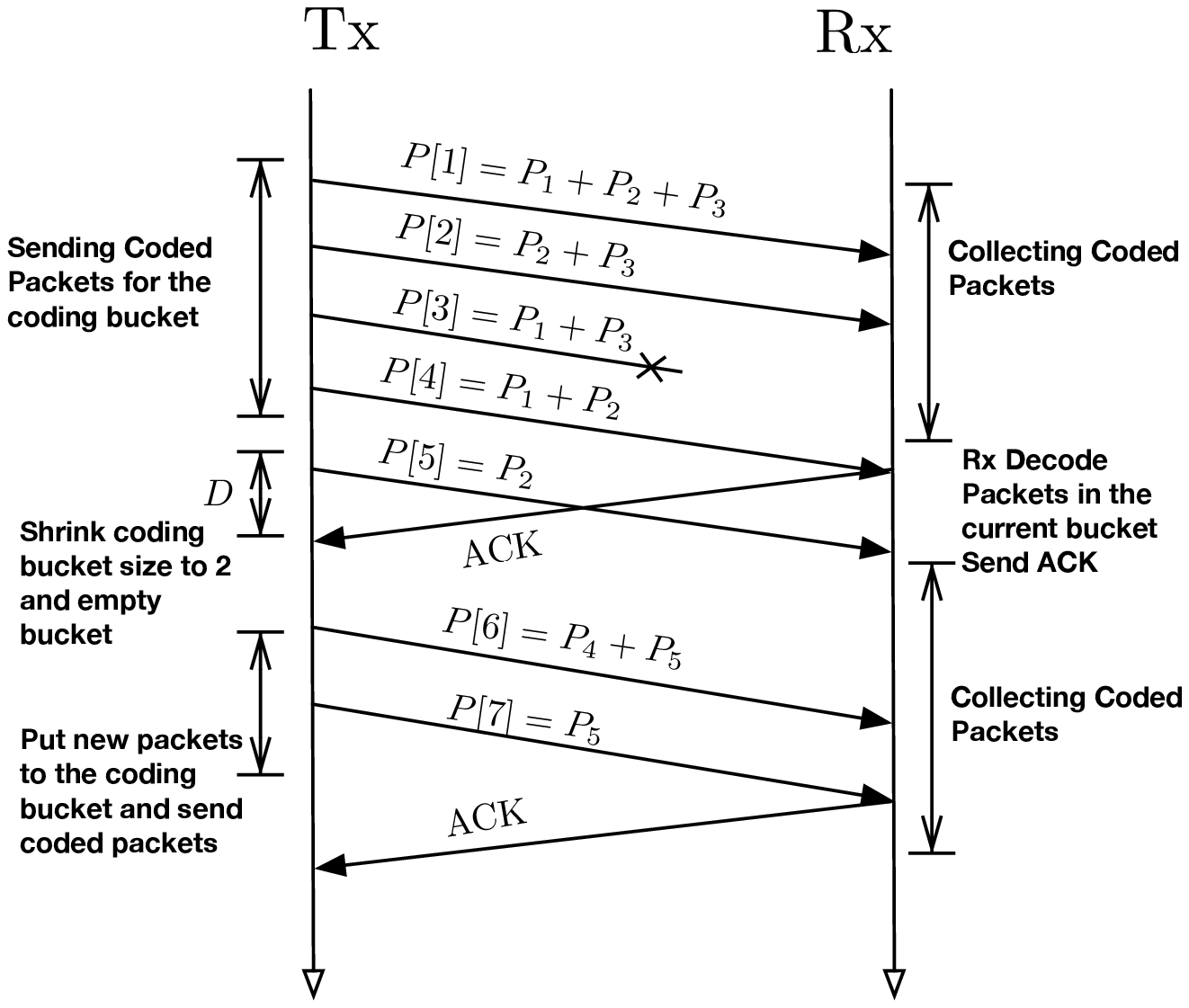}
    \label{fig:codingScheme2}
    }
  \end{center}
  \vspace{-0.3cm}
  \caption{Adaptive Linear Coding Scheme in Point-to-Point Case}
  \label{fig:codingScheme}
\end{figure*}

\begin{figure}[bt]
  \begin{center}
    \includegraphics[scale=0.5]{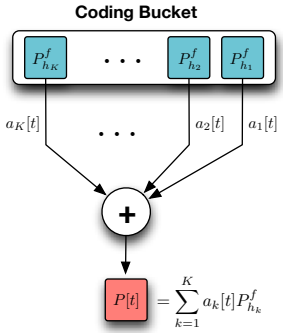}
  \end{center}
  \vspace{-0.5cm}
  \caption{An illustration of encoding}
  \label{fig:encoding}
  \vspace{-0.5cm}
\end{figure}

Consider a point-to-point communication system illustrated in Figure
\ref{fig:codingScheme1}. The sender (Tx) and the receiver (Rx) are connected by
a wireless erasure channel with packet erasure probability $\varepsilon$ and a
perfect feedback channel with delay $D$. The sender looks to transmit to the
receiver a flow $f$ consisting of $N$ packets. The packets are denoted as
$\{P^f_1, \cdots, P^f_N \}$. Each of them is treated as a length $m$ vector in
the space $\mathbb{F}_q^m$, over some finite field $\mathbb{F}_q$. All $N$ data
packets are assumed to be available at the sender prior to any transmissions. In
a fixed generation-based linear network coding scheme, the sender chooses an
integer $K \geq 1$, and sequentially partition the $N$ packets into $\lceil
\frac{N}{K} \rceil$ generations $\{G^f_1, \cdots, G^f_{\lceil
\frac{N}{K}\rceil}\}$, where $G^f_i =
\{P^f_{iK+1},\cdots,P^f_{\min((i+1)K,N)}\}$. At each time slot $t$, the sender
reads the head of the line (HOL) generation $G^f_h = \{P^f_{h_1},\cdots,
P^f_{h_K}\}$, where $h = 1,\cdots, \lceil\frac{N}{K}\rceil$ is the generation
index, and $h_k, k=1,\cdots, K$ are the indices of packets within the
generation.  It then generates a coded packet $P[t]$ that is a linear
combination of all packets in $G^f_h$ (shown in Figure \ref{fig:encoding}), i.e.
\begin{equation}
P[t] = \sum_{k = 1}^{K} a_k[t] P^f_{h_k},
\end{equation}
where $a[t]= (a_1[t],\cdots, a_K[t])$ is the coding coefficient vector, which is
uniformly chosen at random from $\mathbb{F}_q^K$ \cite{ho_random_2006}. The
coded packet, with the coefficient vector appended in the header, is then sent
to the receiver through the erasure channel. The receiver collects coded packets
over time.  Given a large enough field $\mathbb{F}_q$, the receiver, with high
probability \cite{ho_random_2006}, is able to decode the $K$ packets in the
generation through Gaussian elimination performed on the linear system formed on
any $K$ coded packets. Once the receiver decodes the HOL generation
successfully, it sends an ACK message through the feedback channel to the sender.
The sender, who receives the ACK after a delay of $D$ time slots, will purge the
old HOL generation and move on to the next generation in the line.

Our scheme modifies such generation-based network coding in the following ways.
The packets are not partitioned into generations prior to transmission.
Instead, a \textit{coding bucket} is created and acts like the HOL generation.
We use the term \textit{bucket} to avoid confusion with normal generation-based
schemes. The size of the bucket in term of number of packets is denoted as $K$.
The sender collects information about user-end delay constraints and chooses the
bucket size $K$ dynamically. Figure \ref{fig:codingScheme2} gives a simple
example. At the beginning, the coding bucket contains three packets $\{P_1, P_2,
P_3\}$. The sender keeps transmitting encoded packets, i.e. $P[1]$ to $P[5]$, of
these three packets. Upon receiving the ACK feedback, it empties the bucket and
decides to shrink bucket size to $2$, possibly because of the tighter delay
constraint experienced at the receiver. Therefore, only two packets $\{P_4,
P_5\}$ go into the bucket for subsequent transmissions. We leave the details of
adaptively determining coding bucket size to Section \ref{sec:singleCase}. 
\subsection{$\ell_p$-Norm Delay Metrics}
Now we define the delay metrics used in the paper. Following the notations used
in the previous part, let $T_i$ be the time slot in which the packet $P^f_i$ is
decoded at the receiver, and is delivered to upper layer. We require the delivery
of original data packet $\{P^f_1, \cdots, P^f_N\}$ to be \textit{in order}. In the case
when the sequence of packet decoded is out-of-order, we assume that they are
buffered at the receiver to ensure in-order final delivery. $T_i$ represents the
final in-order delivery times of packets $P^f_i$, and we have $T_1 \leq T_2 \leq
\cdots \leq T_N$. We define the inter-arrival times $\Delta T_i$ of the original
packets to be: 
\begin{align}
\Delta T_1 &\triangleq T_1 + D \\
\Delta T_i &\triangleq T_i - T_{i-1}, \quad i = 2, \cdots, N,
     \label{eqn:interdecodingtime}
\end{align}
where $D$ is the feedback delay from the receiver to the sender.  Note that a
feedback message $ACK$ is always assumed to be received correctly after $D$ time
slots. However, when there is more than one receiver, we assume that, in
general, receivers experience different feedback delays across the system owing to
its location and channel variations. Let the size of each data packet be $L$. We
define the delay cost function as a metric of the following form, 
\begin{equation}
  \displaystyle
  d(p) \triangleq \frac{1}{L}\left(\frac{\sum_{i=1}^N(
  \mathbf{E}[\Delta T_i])^p}{N}\right)^{1/p}, \quad p\in [1, \infty),
  \label{eqn:defDelay}
\end{equation}
where $\mathbf{E}[\Delta T_i]$ is the expected value of $\Delta T_i$. The
expectation is taken over the distribution of packet erasures over the system and
all the randomness associated with the coding and scheduling scheme, which are
specified in Section \ref{sec:singleCase}.

Mathematically, the delay metric is a normalized $\ell_p$-norm of the vector
$[\mathbf{E}[\Delta T_1] \cdots \mathbf{E}[\Delta T_N]]^T$. Physically, however,
$p$ measures the delay sensitivity of the receiver and is predominantly
dependent on the type of applications running on the receiver. As the
value of $p$ varies from $1$ to $\infty$, the delay function becomes
increasingly biased towards the large components in the vector, hence indicating
increasing user sensitivity toward large inter-packet delay. As an example,
consider the case when $p=1$.  Since $\sum_{i=1}^N \mathbf{E}[\Delta T_i] =
\mathbf{E}[T_N] + D$, the delay in \eqref{eqn:defDelay} simplifies to,
\begin{equation}
  d(1) = \frac{\mathbf{E}[T_N]+D}{LN},
  \label{eqn:d1delay}
\end{equation}
that is, $d(1)$ is the average delay per packet, normalized by the total size
of the received data. Minimizing $d(1)$, therefore, is equivalent to average
rate maximization for the receiver. On the other hand, consider the case when
$p=\infty$. Because of the max norm, the delay function in \eqref{eqn:defDelay}
reduces to, 
\begin{equation}
  d(\infty) = \frac{\max_i \mathbf{E}[\Delta T_i]}{L}.
  \label{eqn:dmaxdelay}
\end{equation}
Effectively, minimizing $d(\infty)$ translates into minimizing the maximum
expected inter-arrival time between any two successive packets. We call this the
\textit{per-packet delay}.

The flexibility in choosing various $p$-value for delay metrics provides a
unified way of looking at the delay sensitivity at the user side. If a user is
downloading a file, he is certainly more concerned about shortening the overall
completion time or average delay per packet. Consequently, $d(1)$ is the
appropriate delay metric to be optimized.  On the other hand, if the user is
running a real-time video applications, then $d(\infty)$ is more likely to be
the right metric to be minimized as it allows sequence of packets to catch up
quickly with respective delivery deadlines. 
\subsection{Delay In Adaptive Coding Scheme}
In the adaptive coding scheme, a receiver will decode all packets in the current
bucket before informing the sender to empty the bucket and move in new packets.
Assume that the rate at which the coded packets are transmitted is $r$. Consider
the transmission of a bucket of $K$ packets $\{P_{i_1}, \cdots , P_{i_K}\}$.
Once the receiver collects $K$ linearly independent coded packets of the bucket,
it decodes all $K$ packets together. Hence, the ordered inter-arrival times of
original packets will satisfy, $\mathbf{E}[ \Delta T_{i_1} ] = \frac{K}{r} + D$
and $\Delta T_{i_1} = \cdots \Delta T_{i_K} = 0$. In general, consider the case
when the bucket size remains the same for a sequence of $N$ packets,
$\{P_{i_1},\cdots,P_{i_N}\}$. $N$ is divisible by $K$, as the bucket size may
only change when the bucket is emptied. The packets will sequentially enter the
bucket in groups of $K$ packets. Then, for the inter-arrival time of the $j$-th
packet, we have,
\begin{equation}
    \mathbf{E}[\Delta T_{i_j}] = \left\{
  \begin{array}{ll}
    \frac{K}{r} + D, & \text{if } j \equiv 1\pmod{K},  \\
    0, & \text{otherwise}.\\
  \end{array}
    \right.
\end{equation}
Therefore, if the adaptive scheme chooses bucket size of $K$ of a sequence of
$N$ packets, we can simplify \eqref{eqn:defDelay} to measure the delay
cost function for the transmission of the $N$ packets, resulting in: 
\begin{align}
  d(p) &= \frac{1}{L}\left( \frac{\frac{N}{K}\sum_{j=1}^{K} ( \mathbf{E}[\Delta
  T_{i_j}] )^p}{N} \right)^{1/p}  \\
  &= \frac{1}{L}\left( \frac{\frac{N}{K} (\frac{K}{r}+D)^p}{N} \right)^{1/p}  \\
  &= \frac{\frac{K}{r}+D}{LK^{1/p}}.
\end{align}
In particular, under this coding scheme, the delay $d(p)$ seen by the receiver
over the period is independent of $N$ as long as the coding bucket size remains
to be $K$. Hence, we drop $N$ and only consider the bucket size $K$ for rest of
the paper. Furthermore, in practice, $K$ takes only positive integer values in
$[1, K_{max}]$, where $K_{max}$ is the maximum bucket size, limited by the
maximum tolerable computation complexity of the target system. In this work, for
simplicity, we assume that $K$ takes on real value in the same region $[1,
K_{max}]$.

\section{Geometric Programming}
\label{sec:gp}
We give a concise primer of Geometric Programming before looking specifically
into our system model. For more comprehensive coverage of the topic, we refer
the reader to \cite{boyd_tutorial_2007},\cite{chiang_geometric_2005}.  Geometric
program (GP) is a class of mathematical optimization problems characterized by
some special forms of objective functions and constraints. A typical GP is
nonlinear and non-convex, but can be converted into a convex program so that a
local optimum is also a global optimum. The theory of GP has been well studied
since the 60s \cite{duffin_geometric_1967}.  Well developed solution techniques,
such as \textit{interior point methods} are capable of solving GPs efficiently
even for large scale problems.  Many high-quality GP solvers are available
(\text{e.g.} MOSEK and CVX \cite{grant_cvx:_2008}) for providing robust
numerical solutions for generalized GPs (GGP). 

% GP or GGP sees a lot of applications in engineering design and analysis, as they
% can be used to model or approximate a wide variety of practical problems,
% especially in areas such as digital circuit design. In wireless communication,
% GP is often used to solve the transmission power control problems
% \cite{chiang_geometric_2005}. 

Consider a vector of decision variables $\mathbf{x} = [x_1
\dots x_n]^T$. A real function $g: \mathbf{R}^n \rightarrow \mathbf{R}$ is said
to be a \textit{monomial} if it can be written in the form, 
\begin{equation}
  g(\mathbf{x}) = c \prod_{i=1}^{n} x_i^{a_i},
  \label{eqn:monomial}
\end{equation}
where the coefficient $c$ is \textit{positive}, and the exponents $a_1,\dots,
a_n$ are arbitrary real numbers. A function $f: \mathbf{R}^n \rightarrow
\mathbf{R}$ in the form
\begin{equation}
  f(\mathbf{x}) = \sum_{k=1}^{K} c_k \prod_{i=1}^{n} x_i^{a_{ik}},
  \label{posynomial}
\end{equation}
with all $c_k$ being positive real numbers, is called a posynomial. A
posynomial is the sum of arbitrary number of monomials. On top of this, any
function $\tilde{f}$, which can be constructed with posynomials using addition,
multiplication, positive power and maximum operations is called a
\textit{generalized posynomial}. 

A \textit{standard form} geometric program is presented as follows, 
\begin{equation}
\begin{aligned}
  &\text{minimize}  && f_0(\mathbf{x}) \\
  &\text{subject to} && f_i(\mathbf{x}) \leq 1, \; i = 1,\dots,m,
  \label{eqn:standardGP}\\
          &   && g_j(\mathbf{x}) = 1, \; j = 1,\dots,p,  \\
\end{aligned}
\end{equation}
where $f_i(x)$ are posynomials and $g_i(x)$ are monomials, and $x_i$ are the
decision variables, which are also implicitly assumed to be positive, i.e. $x_i
> 0,\; i=1, \dots, n$. In particular, the objective of the optimization has to
be minimizing some posynomial. That says, for solving maximization problems
with GP, the objective function has to be in the form of some monomial
$g(\mathbf{x})$, so that instead of maximizing $g(\mathbf{x})$, we can minimize
$\frac{1}{g(\mathbf{x})}$, which is itself a monomial. In the case where any
$f_i(\mathbf{x})$ is a generalized posynomial, the optimization program is said
to be a \textit{generalized geometric program} (GGP). All generalized geometric
programs can be converted into standard geometric programs and solved
efficiently.

Note that a GP in its standard form is non-convex, as in general posynomials are
non-convex. In order to apply general convex optimization methods, a GP is
usually transformed into its convex form. Let $y_i = \log x_i$ so that $x_i =
e^{y_i}$, the standard form GP can be transformed into its equivalent convex
form, 
\begin{equation}
\begin{aligned}
  &\text{minimize}  && \log f_0(e^{\mathbf{y}}) \\
  &\text{subject to} && \log f_i(e^{\mathbf{y}}) \leq 1, \; i = 1,\dots,m,
  \label{eqn:convexGP}\\
          &   && \log g_j(e^{ \mathbf{y} }) = 1, \; j = 1,\dots,p.  \\
\end{aligned}
\end{equation}
In particular, a monomial constraints 
\[
g_j(\mathbf{x}) = d_j
\prod_{k=1}^{n}x_k^{a_{jk}} = 1
\]is converted to 
\begin{equation}
  \log g_j(e^{\mathbf{y}}) = \log d_j + \sum_{k=1}^{n} a_{ik}\; y_k = 0,
  \label{eqn:logmonomial}
\end{equation}
which is affine and convex. On the other hand, the posynomial parts are
converted into log-sum-exp functions, which can be easily shown to be convex.
Therefore, although the original standard formulations of GPs are nonlinear and
non-convex, they can be converted into convex form as in \eqref{eqn:convexGP}
and solved efficiently. In this paper, we use GP to optimize the coding
parameters and resource allocation at the transmitter with respect to the
$\ell_p$-norm delay metric defined previously.

\section{Single Broadcast Channel With Packet Erasures}
\label{sec:singleCase}
\subsection{System Model}
The motivating scenario of the work comes from a typical home network
environment with multiple user networking devices. The receivers are wirelessly
connected to a WiFi access point (AP), which is then linked with the gateway to
the Internet. All the flow of packets from the Internet to the user devices goes
through the gateway and the access point. The applications running on different
devices have very different delay sensitivities and constraints, as discussed
before. The gateway and the access point look for the optimal coding and
scheduling parameters to ensure the QoE of all the users within the network.

\begin{figure}[ht]
  \begin{center}
    \includegraphics[width=0.45\textwidth]{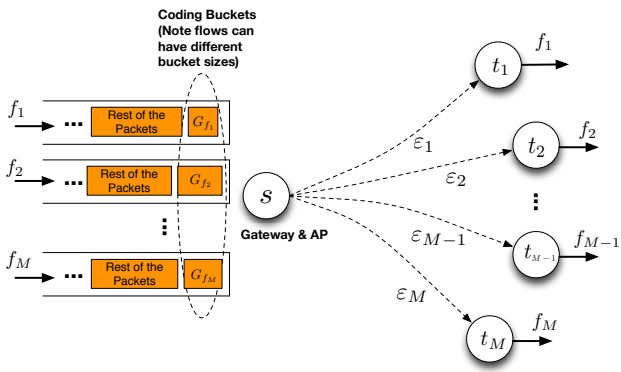}
  \end{center}
  \vspace{-0.2cm}
  \caption{System Model with Single Transmitter}
  \label{fig:system-model}
  \vspace{-0.2cm}
\end{figure}

Conceptually, we represent the system using the following model. We assume that
the link between the AP and gateway has a high capacity and is lossless. Thus,
we represent both the gateway and the AP together as a single node $s$. We
denote the set of receivers by $T=\{t_1,\cdots, t_M\}$. Each receiver needs to
obtain a flow of packets from some source over the Internet. Let $\mathcal{F} =
\{f_1, \cdots, f_M\}$ be the set of flows, where $f_i$ is the packet flow
requested by receiver $t_i$. Note that all $f_i$ enter the system from node $s$,
which in turn acts as a source node. The flows for different receivers are
assumed to be independent. The original data packets in each flow are numbered,
with $P^{f_i}_j$ representing the $j$-th packet in flow $f_i$.  We assume that
there are always enough packets to be served for each flow, since that is the
case when there is a heavy traffic condition. Furthermore, all packets are
assumed to have the same size $L$ in the system and the system is time slotted.
At any time slot, the node $s$ is able to broadcast a size $L$ packet to all
receivers, through the packet erasure broadcast channel. Erasures happen
independently across all receivers and all time slots, i.e. the channel is
memoryless.  We denote the erasure vector by $\mathbf{e} = [\varepsilon_1 \cdots
\varepsilon_M]$, where $\varepsilon_i$ represents the erasure probability seen
by receiver $t_j$. Figure \ref{fig:system-model} gives an illustration of the
system model in the discussion.
\subsubsection{Scheduling Strategies}
Most of the works we discussed in Section \ref{sec:intro} focus on linear
network codes for multicast, in which all the receivers request the same content
from the sources. In the system we consider here, however, we have a multiple
unicast scenario, as each sink looks to receive its own flow, independently from
others. The resource at node $s$ has to be shared among all the receivers. 
Specifically, for every time slot, the sender $s$ has to make a decision on
which receiver to transmit to. While many sophisticated scheduling algorithms
are available, for simplicity, we use a simple stochastic scheduling algorithm.
At any time slot, the node $s$ serves receiver $t_j$ or flow $f_j$ with
probability $a_{j}$, independently from of any other time slots. In the long run,
equivalently, the transmitter node $s$ is spending $a_{j}$ portion of time
serving receiver $t_j$. We call the vector $\mathbf{a} = (a_1,\cdots, a_M)$ the
vector of \textit{scheduling coefficients}.
\subsubsection{Intra-session Coding}
The adaptive coding scheme described in Section \ref{sec:delay} is used in the
system. In the multiple receiver case, we use intra-flow coding, i.e. each
unicast flow is coded independently and separately from others. The coding
bucket sizes and scheduling coefficients, however, are determined by solving
system-wise optimizations. In this case, for a given time slot, if the
transmitter decides to serve receiver $t_j$, it looks for packets in the coding
bucket of flow $f_j$, and encodes these packets using random linear network
codes. The coded packet is broadcasted to all the receivers. With
probability $1-\varepsilon_j$, the targeted receiver $t_j$ will receive it
correctly. Note that we assume the coding coefficients are embedded in the
header of the packet and the size is negligible compared to the size of the
packet $L$. The coding bucket size for flow $f_j$ is denoted as $K_j$. In
general, $K_i \neq K_j$ for $i\neq j$, and $K_j$ may vary over time as the delay
requirements at the receivers changes. Let $\mathbf{K} = (K_1,\cdots, K_M)$. We
aim to optimize both $\mathbf{a}$ and $\mathbf{K}$, based on the varying delay
constraints at the receivers.
\subsection{Delay Optimization}
We first consider the case where there is only a single receiver, i.e.
$M = 1$. Since there is no scheduling issue or system-wise fairness
consideration in the case, it makes sense to minimize the delay cost
function associated with the receiver. As there is no ambiguity of
notations, we drop all the subscripts. It is easy to see that the packet
transmission rate in this case is $1-\varepsilon$ for the receiver and thus the
expected time for receiving $K$ coded packets is $\frac{K}{1-\varepsilon}$.
Subsequently, the $\ell_p$-norm delay cost function minimization problem is
given as follows,
\begin{align}
  \text{minimize}&  \quad d(p) = \frac{\frac{K}{1-\varepsilon}+D}{LK^{1/p}} 
  \label{eqn:delayOpt} \\
  \text{subject to}& \quad  1 \leq K \leq K_{max}.
\end{align}
The optimal block size $K^*$ can be obtained by setting zero to the gradient of
the Lagrangian of objective function. We have 
\begin{equation}
  K^* = \left(\frac{(1-\varepsilon) D}{p-1}\right)_{[1,K_{max}]}, \quad 0 <
  \varepsilon < 1,
  \label{eqn:optK}
\end{equation}
where the subscript denotes the projection, 
\[
(x)_{[a,b]} \triangleq
\min(\max(a,x),b).
\]
For better understanding of the delay metrics, consider the relation between
$d(1)$ and $d(\infty)$. From \eqref{eqn:defDelay}, we have,
\begin{equation}
K =
\frac{D(1-\varepsilon)}{(1-\varepsilon)Ld(1) -1}.
\end{equation}
Hence, the trade-off between
$d(1)$ and $d(\infty)$ can be expressed as follows,
\begin{equation}
  d(\infty) = \frac{D}{L-\frac{1}{d(1) (1-\varepsilon)}}.
  \label{eqn:d1df}
\end{equation}
Ignoring the bucket size constraints for simplicity, given $D$, we can vary $K$
from $1$ to $\infty$, and plot the values of $d(\infty)$ against $d(1)$ for the
trade-off curve. Each point on the curve corresponds to a choice of $K$, which
is equivalent to a choice of optimizing $d(p)$ for some $p$, because of
\eqref{eqn:optK}. Therefore, the choice of $p$ at the receiver indicates the
a point on the trade-off curve of $d(1)$ and $d(\infty)$ that is desired by the
receiver.

We can also use the zero duality gap in GP to obtain the optimal $d(p)$ directly
from the dual function. Note that the dual function of \eqref{eqn:delayOpt} is
given by,
\begin{equation}
  v =
  \left(\frac{1}{(1-\varepsilon)L\beta_1}\right)^{\beta_1}\left(\frac{D}{L\beta_2}\right)^{\beta_2}
  \label{eqn:dualopt}
\end{equation}
where $\mathbf{\beta}=(\beta_1,\beta_2)$, can be obtained from solving a simple
linear system,
\begin{equation}
\begin{cases}
  (1-1/p) \beta_1 + (-1/p) \beta_2 &= 0, \quad (\text{normality condition}) \\
  \beta_1 + \beta_2 &= 1, \quad (\text{orthogonality condition})
\end{cases}
\end{equation}
\subsection{Delay Constrained Optimization with GP}
\subsubsection{GP Formulation}
For $M > 1$, instead of minimizing the delay of a specific receiver, we are
interested in optimizing certain system-wise utility function with the
constraints that the $\ell_p$-norm delay requirements must be satisfied at each
receiver. We assume the each receiver $t_j$ monitors the delay constraints for
targeted QoE of its applications and set a maximum acceptable delay
$\hat{d}(p_j)$, corresponding to its delay sensitivity $p_j$. For the objective
function, we choose to maximize the min rate of all receivers. If the packet
transmission rate to $t_j$ is $r_j$, then the actual data rate received by $t_j$
is $\frac{LK_j}{\frac{K_j}{r_j}+D_j}$, where $D_j$ is the feedback delay of
$t_j$. Let $\mathbf{r}=(r_1,\cdots,r_M)$. The optimization problem is then given
as follows: 
\begin{align}
  \text{max }_{\mathbf{K}, \mathbf{r}, \mathbf{a}} \quad &\min_{j} \frac{LK_j}{\frac{K_j}{r_j}+D_j} \\
    \text{subject to} \quad & \frac{\frac{K_j}{r_j} + D_j}{L
    K_j^{1/p_j}} \leq \hat{d_j}(p_j) && \forall j=1,\dots,M
    \label{eqn:delayCons-single}\\ 
    & r_j \leq a_j(1-\varepsilon_j) && \forall j=1,\dots,M
    \label{eqn:rateCons-single}\\
    & \sum_{j} a_{j} \leq 1 && \label{eqn:schedulingCon-single} \\
    & 1 \leq K_j \leq K_{max} && \forall j = 1, \dots, M.
    \label{eqn:complexityCon-single}
\end{align}
In the above formulation, constraints \eqref{eqn:delayCons-single} and
\eqref{eqn:rateCons-single} represent the delay and rate constraints
respectively for receiver $t_j$, while \eqref{eqn:schedulingCon-single} is the
scheduling probability constraint at the sender node $s$. The problem is a
Generalized Geometric Program. In particular, all constraints can be converted
into upper bound of posynomials of $\mathbf{K}, \mathbf{r}$ and $\mathbf{a}$.
The only non-posynomial part is the objective function, which can be transformed
into upper bounding posynomial constraints and monomial objective by adding
auxiliary variable $x$, 
\begin{align}
  \max_{\mathbf{K}, \mathbf{r}, \mathbf{a}, x } \quad & x \\
  \text{subject to} \quad & \frac{x(\frac{K_j}{r_j}+D_j)}{L K_j} \leq 1, &&
  \forall j.
\end{align}
Combining this with \eqref{eqn:delayCons-single} to
\eqref{eqn:complexityCon-single}, we have a GP that can be efficiently solved.
\subsection{Illustrations of Trade-offs}
\subsubsection{Trade-off: Average Delay vs Per-packet Delay}
\begin{figure}[bth]
  \centering
    \includegraphics[width=0.40\textwidth]{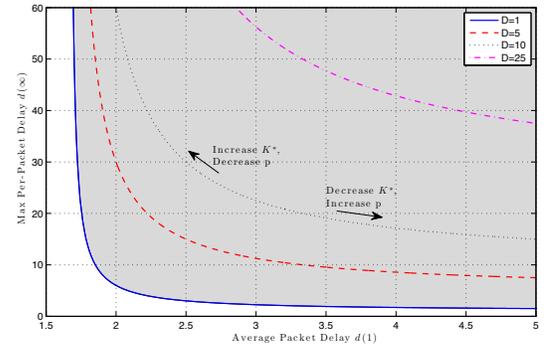}
    \vspace{-0.2cm}
    \caption{Tradeoff of $d(1)$ vs $d(\infty)$ with varying $D$}
  \label{fig:tradeoffDK}
\end{figure}
Figure \ref{fig:tradeoffDK} demonstrate the trade-off between $d(1)$ and
$d(\infty)$ following Equation \eqref{eqn:d1df} with various values of $D$ and
erasure probability $\varepsilon=0.4$. As discussed previously, if we
parameterize $d(1)$ and $d(\infty)$ on the optimal bucket size $K^*$, as $p$
varies from $1$ to $\infty$, we obtain the same curves. The shaded area bounded
by each curve is the area of all achievable pairs $(d(1), d(\infty))$ for the
specific feedback delay.  With small $D$, both low delay in $d(1)$ and
$d(\infty)$ can be achieved.  However, when feedback delay increases, the
trade-off becomes increasingly stronger. This is evident from Equation
\eqref{eqn:d1df}, where $D$ appears in the numerator. It is expected as for
average delay, coding over larger generations amortizes the feedback delay over
more packets.  But for the per-packet delay $d(\infty)$, increased feedback
delay must be compensated by even smaller generation size for more frequent
decoding. This is also consistent with Equation \eqref{eqn:optK} where $K^*$
increases with feedback delay $D$ and decreases as delay sensitivity $p$.

\subsubsection{Adaptive Scheme vs Fixed Generation Coding}
\begin{figure}[bt]
  \centering
    \includegraphics[width=0.40\textwidth]{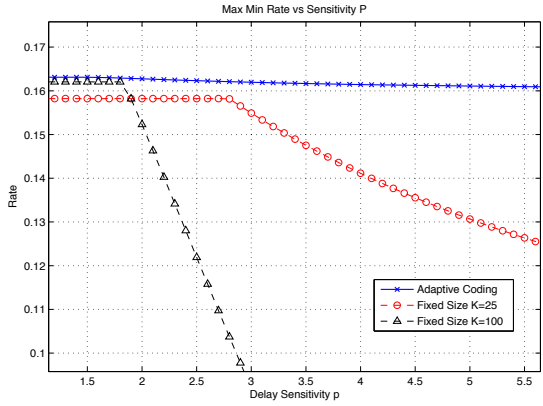}
    \vspace{-0.2cm}
    \caption{Min Rate vs Delay Sensitivity $p_1$}
    \label{fig:maxMinRate}
\end{figure}

\begin{figure}[bt]
  \centering
    \includegraphics[width=0.40\textwidth]{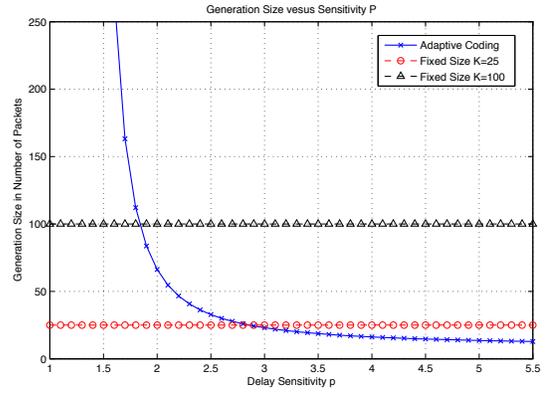}
    \vspace{-0.2cm}
    \caption{Coding Bucket Size vs Delay Sensitivity $p_1$}
    \label{fig:genSize}
\end{figure}

\begin{figure}[bt]
  \centering
    \includegraphics[width=0.40\textwidth]{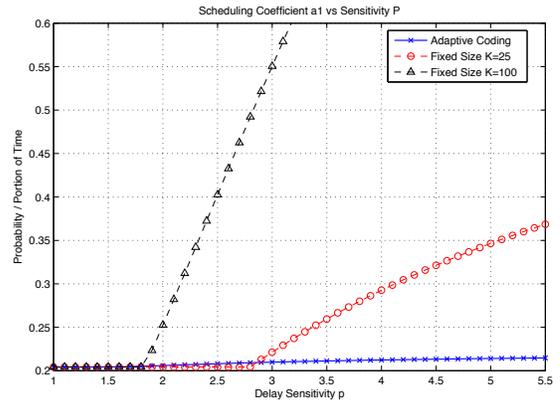}
    \vspace{-0.2cm}
    \caption{$a_1$ vs Delay Sensitivity $p_1$}
    \label{fig:a1vsP}
\end{figure}

Figure \ref{fig:maxMinRate} to \ref{fig:a1vsP} shows some comparisons between
adaptive coding schemes with fixed generation size coding schemes, as the delay
sensitivity $p_1$ of the first receiver increases. In this example, we have $5$
receivers, with erasure $\mathbf{e}=[0.4, 0.1, 0.15, 0.2, 0.25]$, the same
$D=5$, $L=1$ and $\hat{d}_j = 50/L$. Except for receiver $1$, whose $p_1$ value
varies, we have $p_j=1$ for all other receivers. For the fixed generation size
schemes, we choose $K=25$ and $K=100$ for representing small and large
generation respectively. In this cases, scheduling coefficients are the only
decision variables in the optimization for min rate. From Figures
\ref{fig:maxMinRate} and \ref{fig:genSize}, we can see that, initially, the min
rates for different schemes are relatively close. The adaptive scheme is able
to choose a much larger coding bucket size to obtain some rate gain compared to
$K=25$ case. As $p_1$ increases, the fixed coding generation schemes are unable
to reduce generation size. In order to meet the growingly stringent delay
constraint, the sender has to devote increasingly more time to receiver $1$,
as seen in Figure \ref{fig:a1vsP}. Inevitably, the time for serving other
receivers is greatly reduced and the min rate of the system decreases quickly.
In the $K=100$ case, the delay requirements cannot be satisfied for $p_1 > 2.9$.
On the contrary, for the adaptive scheme, which optimizes bucket size and
scheduling jointly, there is little decrease in min rate. For low delay
sensitive receivers, the scheme will assign them large coding bucket sizes to allow
rate gain. As a results, the sender is able to meet their delay-rate constraints
with less serving time and save time for higher receivers. On the other hand, as
$p$ values for some receiver increases, its coding bucket size is reduced to
quickly decreases the per-packet delay. Hence, the scheme is able to accommodate
high delay sensitive receivers much better. 

\section{Multiple Wireless Packet Erasure Channels}
\label{sec:multiCase}
\begin{figure}[bt]
  \begin{center}
    \includegraphics[width=0.35\textwidth]{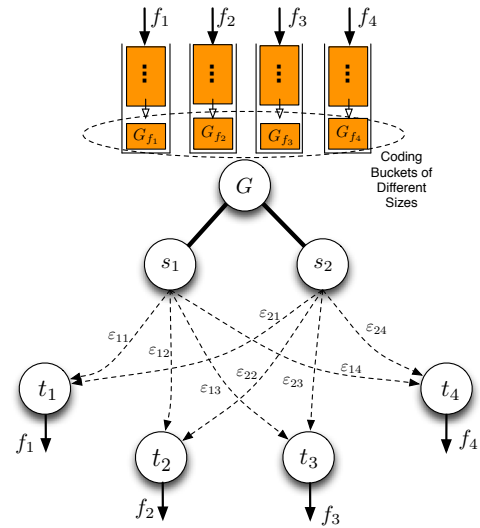}
  \end{center}
  \vspace{-0.2cm}
  \caption{An example system with $2$ senders and $4$ receivers}
  \vspace{-0.2cm}
  \label{fig:systemMultiAPs}
\end{figure}
With the proliferation of low cost access points, many devices may be covered by
more than one access points in wireless home, campus or enterprise networks.
That leads to an important extension of the work to the case of multiple
broadcast erasure channels covering the same set of receivers. As in the
previous section, we still have the same set of receivers, $T=\{t_1, \cdots,
t_M\}$. However, there are now $W$ access points, or transmitters, denoted by
the set $S=\{s_1,\cdots,s_W\}$.  Instead of an erasure probability vector, we
have an erasure probability matrix $\mathbf{e}=[\varepsilon_{ij}]$, where
$\varepsilon_{ij}$ is the erasure probability between node $s_i$ and $t_j$. An
example of the system is illustrated in Figure \ref{fig:systemMultiAPs}. We
assume that the channels are orthogonal or non-interfering. 

The same coding and scheduling scheme is used for the new system. We use
$\mathbf{a}=[a_{ij}]$ to represent the probability of transmitter $s_i$ serving
the flow $f_j$ at any time slot. The scheduling and coding optimization is done
at the gateway node $G$, who coordinates all senders who perform encoding.
Furthermore, for each flow $f_j$ in $\mathcal{F}$, all senders in $S$ have the
same coding bucket size $K_j$, dictated by node $G$. This ensures that, for each
flow, every sender sends coded packets in the coding buckets consisting of the same
data packets and guarantees the decodability. An important feature of the
randomly linear coding is that all coded packets from the same bucket are
exchangeable. That avoids complicated scheduling based on sequence numbers of
the uncoded packets and helps to reduce transmission redundancy in erasure
channels.

\subsection{Signomial Program Formulation}
Similarly to the single sender case, we can formulate an optimization program for
determining $\mathbf{K}, \mathbf{a}$ and $\mathbf{r}$. For example, the rate
product maximization is given as, 
\begin{align}
  \min_{\mathbf{K}, \mathbf{r}, \mathbf{a}, \mathbf{R}} \;&\prod_{j} R_j^{-1} \\
    \text{subject to} \;& \frac{\frac{K_j}{r_j} + D_j}{L
    K_j^{1/p_j}} \leq \hat{d_j} && \forall j=1,\dots,M\label{eqn:delayCons-multi}\\ 
    & r_j \leq \sum_i a_{ij}(1-\varepsilon_{ij}) && \forall j=1,\dots,M
    \label{eqn:rateCons-multi} \\
    & R_j \leq \frac{LK_j}{\frac{K_j}{r_j}+D_j} && \forall j=1,\dots,M
    \label{eqn:avgRCons-multi}\\
    & 1 \leq K_j \leq K_{max} && \forall j = 1, \dots, M
    \label{eqn:complexityCons-multi}\\
    & \sum_{j} a_{ij} \leq 1 &&  \forall i = 1, \dots, W,
    \label{eqn:schedulingCons-multi}
\end{align}
where the delay constraint \eqref{eqn:delayCons-multi} and complexity constraint
\eqref{eqn:complexityCons-multi} remain the same. Auxiliary variables $R_j$ and
\eqref{eqn:avgRCons-multi} are used to represent the average rates for the
receiver.  Maximizing rate product is equivalent to minimizing the product of
average delays $R_j^{-1}$, hence the objective $\prod_j R_j^{-1}$. The packet
transmission rate for each receiver in this case is bounded by $\sum_i
a_{ij}(1-\varepsilon_{ij})$. However, owing to the existence of this new
transmission rate constraint \eqref{eqn:rateCons-multi}, the problem becomes
truly non-convex. In particular, the constraint can be written as 
\begin{equation}
  r_j + \sum_i (-a_{ij})(1-\varepsilon_{ij}) \leq 0, 
  \label{eqn:rateCons-signomial}
\end{equation}
which is an upper bound constraint on a signomial. A signomial is a sum of
monomials whose multiplicative coefficients can be either positive or negative.
The problem therefore belongs to a more general class of problem called
\textit{Signomial Program}, which is truly non-convex and NP-hard in general.
Only local optimal solutions can be efficiently computed. Based on the most
widely used monomial condensation methods, we provide an efficient way to
approximate the solution with successive GP solutions. 

\subsection{Successive GP Approximation}
Consider an arbitrary signomial $h(\mathbf{x})$. It can always be written as
the difference between two posynomials, i.e.  $h(\mathbf{x})=f^+(\mathbf{x}) -
f^-(\mathbf{x})$. The inequality $h(\mathbf{x}) \leq 0$ is then equivalent to
$\frac{f^+(\mathbf{x})}{f^-(\mathbf{x})} \leq 1$. We can approximate the left
hand side with a posynomial using common \textit{condensation methods}
\cite{chiang_geometric_2005}. In \textit{single condensation}, the posynomial
denominator $f^-(\mathbf{x})$ is approximated using a monomial
$g^-(\mathbf{x})$, which in turn allows
$\frac{f^+(\mathbf{x})}{f^-(\mathbf{x})}$ to be approximated by a posynomial
$\frac{f^+(\mathbf{x})}{g^-(\mathbf{x})}$. In \textit{double condensation}, both
$f^+$ and $f^-$ are approximated using monomials, which creates a monomial
approximation of $\frac{f^+(\mathbf{x})}{f^-(\mathbf{x})}$. In our case, both
methods are equivalent, since we have $f^+(\mathbf{x}) = r_j$, which is itself a
monomial.  One of the commonly used condensation methods is based on the
following Lemma \cite{chiang_geometric_2005}. 
\noindent\begin{lemma}
Given a posynomial $f(\mathbf{x})= \sum_i u_i(\mathbf{x})$, choose $\beta_i >
  0$, such that $\sum_i \beta_i = 1$, then the following bound holds,
  \begin{equation}
    f(\mathbf{x}) \geq g(\mathbf{x}) = \prod_i
    \left(\frac{u_i(\mathbf{x})}{\beta_i}\right)^{\beta_i}.
    \label{eqn:AMGM}
  \end{equation}
  Furthermore, equality holds when $\mathbf{x} = \mathbf{x}_0$ and $\beta_i =
  \frac{u_i(\mathbf{x}_0)}{f(\mathbf{x}_0)}$. 
  \label{lma:monoApproxi}
\end{lemma}
\textit{Proof: } The results can be easily proved using Inequality of Arithmetic
and Geometric Mean (AM-GM). 

Using Lemma \ref{lma:monoApproxi}, we can approximate constraint
\eqref{eqn:rateCons-multi} in the signomial program with the following,
\begin{equation}
      r_j \leq \prod_i \left( \frac{a_{ij}(1-\varepsilon_{ij})}{\beta_{ij}}
      \right)^{\beta_{ij}}. 
      \label{eqn:monomialapp}
\end{equation}
In particular, the optimization program is then a Geometric Program, if we
replace \eqref{eqn:rateCons-multi} with \eqref{eqn:monomialapp}.
Furthermore, given the monomial approximation in \eqref{eqn:monomialapp}, we can construct
successive GP based on refined approximations of constraint
\eqref{eqn:rateCons-multi} to approach local optimal solutions of the original
Signomial Problem. The algorithm is summarized in Algorithm \ref{alg:approxiSP}.
\begin{algorithm}
  \caption{Successive GP Approximation of SP\label{alg:approxiSP}}
  \textbf{Begin: } A feasible solution $(\mathbf{K}^0,
  \mathbf{a}^0, \mathbf{r}^0, \mathbf{R}^0)$, $t=0$\; 
  \Repeat{Convergence}{
  Compute $f(\mathbf{a}^t) = \sum_i a_{ij}^t (1-\varepsilon_{ij})$\;
  Compute $\beta_{ij} = \frac{a_{ij}^t (1-\varepsilon_{ij})}{f(\mathbf{a}^t)}$\;
  Construct the $t$-th approximation and replace constraint
  \eqref{eqn:rateCons-multi} with the monomial constraint,
  \[
  r_j \leq g(\mathbf{a^t}) = \prod_i \left( \frac{a_{ij}^t(1-\varepsilon_{ij})}{\beta_{ij}}
      \right)^{\beta_{ij}} \;
  \]
  $t = t+1$\;
  Solve the resulting GP to get $(\mathbf{K}^t,
  \mathbf{a}^t, \mathbf{r}^t, \mathbf{R}^t)$\;
  }
\end{algorithm}
\vspace{-0.3cm}
\subsection{Convergence}
\begin{figure}[ht]
  \begin{center}
    \includegraphics[width=0.40\textwidth]{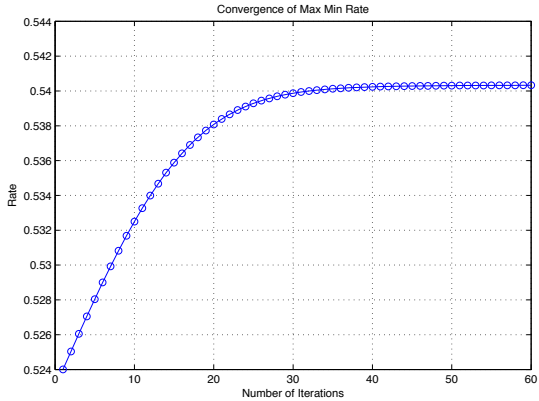}
  \end{center}
  \vspace{-0.5cm}
  \caption{Convergence of Optimal Min rate in the example system}
  \vspace{-0.5cm}
  \label{fig:maxMinConv}
\end{figure}

\begin{figure}[ht]
  \begin{center}
    \includegraphics[width=0.40\textwidth]{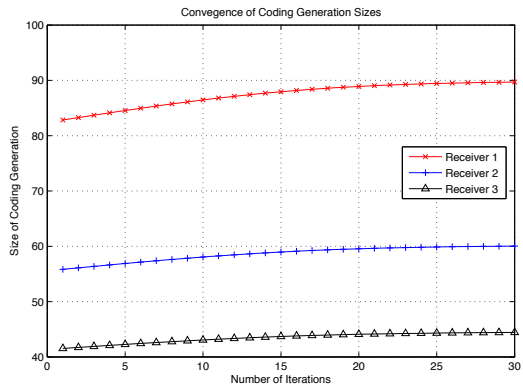}
  \end{center}
  \vspace{-0.4cm}
  \caption{Convergence of Bucket Sizes in the example system}
  \vspace{-0.5cm}
  \label{fig:sizeConv}
\end{figure}

Given Lemma \ref{lma:monoApproxi}, it is easy to show the Algorithm
\ref{alg:approxiSP} always converges.  According to Lemma \ref{lma:monoApproxi},
the values $\beta_{ij}$ are chosen such that for the local approximation at
$\mathbf{a}^t$ in the $t$-th iteration, we have,
\begin{equation}
  g(\mathbf{a}^t) = f(\mathbf{a}^t) \geq f(\mathbf{a}^{t-1}).
\end{equation}
Let the optimal objective for iteration $t$ be $Z^{*,t}$. Then we have $Z^{*,t}
\leq Z^{*,t-1}$. Furthermore, at local optimal $\mathbf{a}^*$, it can be
verified, that $f(\mathbf{a}^*) = g(\mathbf{a}^*)$ and $\nabla f(\mathbf{a}^*) = \nabla
g(\mathbf{a}^*)$, which shows that the algorithm will indeed converge to an
optimal that satisfies the KKT condition. In fact, in many cases, it converges
to the global optimum. Figure \ref{fig:maxMinConv} and \ref{fig:sizeConv} shows
the convergence of min rate and bucket sizes for a example system with $3$
receivers and $2$ transmitters. 
\vspace{-0.3cm}

\section{Conclusions}
\label{sec:conclusion}
In this paper, we consider the trade-off between rate and delay in single-hop
packet erasure broadcast channels with random linear coding schemes. We
characterize the delay and rate requirements of various users with a unified
framework based on the $\ell_p$-norm delay metrics defined on the in-order
packet arrival times. Using the optimal trade-off curve between the average
delay, which can be viewed as the inverse of rate, and the per-packet delay, we
demonstrate how feedback delays and the choice of coding bucket sizes affect the
trade-offs. In the multiple receiver case, we formulate geometric optimization
problems to exploit the trade-off together with the transmission time allocate
at the senders. With an adaptive coding scheme, for low delay sensitive
receiver, the sender could allocation less time while compensating the rate loss
with larger coding bucket. That allows the sender to allocate more time to high
delay sensitive receivers who, at the same time, are assigned with smaller
coding bucket sizes. We show that the adaptive scheme is more robust and
resilient toward high and varying delay sensitivities, since the feedback
information about receiver delay constraints adds extra flexibility to the coding
and scheduling design. In particular, in many systems, this comes with little
cost because of the availability of feedback channels. Finally, when there are
multiple senders, we formulate the same optimization problem into a non-convex
signomial problem and approximate the solution with successive GP approximations
based on single condensation methods and we demonstrate the convergence of the
algorithm.

% needed in second column of first page if using \pubid
%\pubidadjcol

% trigger a \newpage just before the given reference
% number - used to balance the columns on the last page
% adjust value as needed - may need to be readjusted if
% the document is modified later
%\IEEEtriggeratref{8}
% The "triggered" command can be changed if desired:
%\IEEEtriggercmd{\enlargethispage{-5in}}

% references section
\bibliographystyle{IEEEtran}
\bibliography{IEEEabrv,main}
\end{document}